\documentstyle[preprint,prd,aps]{revtex}

\advance\hoffset by -3mm  
\advance\voffset by  8mm  
\def\kp{K^+\to \pi^+\nu\bar{\nu}}
\def\be{\begin{eqnarray}}
\def\ee{\end{eqnarray}}
\def\nn{\nonumber}

\begin{document}

\bibliographystyle{prsty}

\draft

\author{C.Q.~Geng$^1$, I.J.~Hsu$^2$ and Y.C.~Lin$^2$}
\address{${}^1$Physics Department, National Tsing Hua University,
Hsingchu, Taiwan
\\
${}^2$Department of Physics and Astronomy, National Central
University, Chungli, Taiwan}

\title{Comments on long distance contribution to $\kp$}
\maketitle

\narrowtext

\begin{abstract}
We study the long distance contribution to $\kp$ by using chiral
perturbation theory. We find that the tree level $O(P^2)$ contribution
vanishes identically without assuming the large $N_c$ limit. The leading
contribution arises from the one-loop
amplitude, which is $O(P^4)$ in the chiral power counting. The branching
ratio of the long distance contribution is found to be of order $10^{-7}$
smaller compared with that of short distance one and hence is
completely negligible.
\end{abstract}



\newpage
\noindent
{\em Introduction:} The rare decay mode $\kp$ is suppressed by the GIM
mechanism, the leading contribution starts from the one loop level for
the short distance effect. It also receives contribution from the long
distance effect. The heavy quark
effect manifests itself in very different fashions for these two kinds of
contributions. For the long distance effect, it realizes itself as the
coefficients of the low energy effective lagrangian which contains no explicit
heavy quark mass dependence. While for the short distance effect, it appears
explicitly in the propagators of one loop amplitude for the process of
interest. Due to the explicit dependence of heavy quark mass, in particular
$m_t$, the short distance effect dominates the total amplitude. Since the short
distance contribution contains explicit dependence on $m_t$ and $V_{td}$, it
offers an opportunity for experimentalists to measure the standard model
parameters. Because the amplitude is GIM suppressed, it also leaves a window
for physics beyond the standard model.

The short distance contribution has been calculated by Inami and
Lim\cite{inam81} and the long
distance one has been calculated by
Rein and Sehgal\cite{rein89}, by Hagelin and Littenberg\cite{hage89} and
more recently by Lu and Wise\cite{lu94}. It is
the purpose of the present work to reanalyze $\kp$ within the same framework,
namely chiral perturbation theory, as in \cite{lu94}. With different
identification of left-handed and right-handed currents, we find that
the tree level amplitude vanishes identically without assuming the large
$N_c$ limit, as opposed to the result in \cite{lu94}. The leading
contribution arises from one loop amplitude and the branching ratio of
the long distance contribution
is about $10^{-7}$ smaller than that of the short distance one
\cite{litt89,bela91}.
So the $\kp$ mode is virtually a pure short distance effect. This leads
to the conclusion that the accuracy of the determination of the standard
model parameters by extracting from the experimental data is as good as the
experimental measurement.

\noindent
{\em Chiral Lagrangian:} The chiral lagrangian incorporates the external
fields into the covariant derivatives according to the handness of the
external fields. In the standard model, the $Z^0$ acquires both the
left-handed and right-handed components due to the mixing with the
hypercharge. In the three flavour space, the matrix representing the
$Z^0$ particle contains a singlet component $I$, which is not included in
the $SU(3)\times SU(3)$ chiral
symmetry. In order to incorporate the singlet piece into the chiral lagrangian,
we first assume nonet symmetry and then use a nonet symmetry breaking parameter
$\xi$ to indicate the degree of nonet symmetry breaking, $\xi =1$ for exact
nonet symmetry. The covariant derivative is then read
\be
D_\mu U & = & \partial_\mu U-i r_\mu U+i U l_\mu
\nn\\
& = & \partial_\mu U+{i g \over cos \theta_W} (U Q-{\xi\over 6} U I
-sin^2\theta_W \left[U,Q\right])Z_\mu^0,
\ee
where $Q$ is the quark charge matrix, $Q= diag(2/3,-1/3,-1/3)$, and $U$ is the
nonlinear realization of meson octet
\be
U = exp (i \Phi / f_\pi),
\ee
in which
\be
\Phi = \phi^a \lambda^a = \sqrt{2}  \left(
\begin{array}{ccc}
\pi^0/\sqrt{2}+\eta/\sqrt{6}& \pi^+& K^+ \\
\pi^-& -\pi^0/\sqrt{2}+\eta/\sqrt{6}& K^0 \\
K^-& \bar{K}^0& -2\eta/\sqrt{6}
\end{array}
\right)
\ee
and $f_{\pi}= 93 MeV$ is the pion decay constant. Note that this identification
of covariant derivative is different from that in \cite{lu94}.
The covariant derivative in Eq. (16) of {\it Ref.} \cite{lu94}
has a wrong identification of left-handed and right-handed currents,
namely the currents have been placed in wrong positions.\footnote{Our
covariant derivative in Eq. (1) is the same as that, for example,
in Eq. (3.11) of {\it Ref.} \cite{pich95}.}

The chiral lagrangians, in terms of the covariant derivative so constructed,
are given by
\be
{\cal L}_2
& = & {f_{\pi}^2\over 4} Tr \left[ D_{\mu}U^\dagger
D^{\mu}U+2 B_0 M (U+U^\dagger)\right]
\ee
for strong interaction, and
\be
{\cal L}^{\Delta S=1}_2
& = & G_8 f_{\pi}^4 Tr \lambda_6 D_{\mu}U^\dagger
D^{\mu}U
\ee
for weak interaction. The coefficient $G_8$ is related to the Fermi constant
and the CKM matrix elements, the numerical value is given by $G_8=9.1\times
10^{-6}GeV^{-2}$. There are two ways to calculate the tree level amplitude of
$\kp$, using the conventional basis in Eq. (3) or using the
diagonalized basis \be
\pi^+ \to \pi^+ - {2 m_K^2 f_\pi^2 G_8 \over m_K^2 - m_\pi^2} K^+
\ee
\be
K^+ \to K^+ + {2 m_\pi^2 f_\pi^2 G_8^* \over m_K^2 - m_\pi^2} \pi^+
\ee
\be
\pi^0 \to \pi^0 + {\sqrt{2} m_K^2 f_\pi^2 \over m_K^2 - m_\pi^2}
( G_8 K^0 + G_8^* \bar{K}^0)
\ee
\be
K^0 \to K^0 - {\sqrt{2} m_\pi^2 f_\pi^2 G_8^* \over m_K^2 - m_\pi^2} \pi^0
+ \sqrt{{2 \over 3}}{m_\eta^2 f_\pi^2 G_8^* \over m_\eta^2 - m_K^2} \eta
\ee
\be
\eta \to \eta - \sqrt{{2 \over 3}} {m_K^2 f_\pi^2 \over m_\eta^2 - m_K^2}
(G_8 K^0 + G_8^* \bar{K}^0)
\ee
which eliminates the $K^+ - \pi^+$ mixing\cite{ecke88}. There are
three Feynman diagrams, {\it Fig.} 1(a), 1(b) and 1(c),
which contribute to the tree level amplitude in the conventional basis
while, in the diagonalized basis, the amplitude could only receive
a contribution from {\it Fig.} 1(a) since the vertex
$K^+ - \pi^+$ is removed.

We now evaluate the contribution arising from the terms proportional to
$I$, $Q$ and the $[Q,U]$ separately in the conventional basis.
The singlet part, proportional to $\xi$, gives no
contribution to  $K^+ K^+ Z^0$, $\pi^+ \pi^+ Z^0$ and $K^+ \pi^+ Z^0$
vertices and thus it does not appear in the amplitudes.
The rest two parts have nonvanishing contributions to the vertices and,
explicitly, they lead to the following amplitudes
\be
A^{(a)} & = & -{2 i g G_8 f_\pi^2 (1-2\sin^2\theta_W)\over\cos \theta_W}p_K
\cdot \epsilon \nn\\
A^{(b)} & = & -{2 i g G_8 f_\pi^2 (1-2\sin^2 \theta_W)\over\cos \theta_W}
{m_\pi^2 \over m_K^2-m_\pi^2} p_K \cdot \epsilon \nn\\
A^{(c)} & = & {2 i g G_8 f_\pi^2 (1-2\sin^2\theta_W) \over\cos \theta_W}
{m_K^2 \over m_K^2-m_\pi^2} p_K \cdot \epsilon.
\ee
However, the sum of the above three amplitudes in Eq. (11)
is zero, i.e.,
\be
A^{(a)}+A^{(b)}+A^{(c)}=0.
\ee
In the diagonalized basis, it is easy to show that the vertex $K^+\pi^+ Z^0$
arising from the only possibly diagram {\it Fig.} (a) also vanishes,
resulting in a vanishing amplitude. This result differs from
that in {\it Ref.} \cite{lu94} significantly.

\noindent
{\em One-loop Amplitude:} The amplitude receives contribution starting from
$O(P^4)$ in the chiral power counting. In the $O(P^4)$ chiral lagrangian of
weak interaction, the number of counter terms is so large\cite{kamb90} that
fitting the coefficients from
data is beyond the reach of current experiments. As a standard practice, we
only calculate the finite part of the one loop amplitude, the so-called chiral
logarithmic piece, to estimate the order of magnitude of the decay rate.
The amplitude of $\kp$ is found to be
\begin{equation}
A(\kp)  = \nn\\
-{i\alpha G_8 (1-2\sin^2 \theta_W)\over 64\pi M_Z^2\sin^2 \theta_W\cos^2
\theta_W} J(m_K^2) (P_K + P_\pi)^{\mu}\bar{\nu}\gamma_{\mu}(1-\gamma_5)\nu ,
\end{equation}
where the loop function is defined as
\be
J(m^2) & = & {1 \over i\pi^2} \int d^nq {1\over q^2 - m^2}\nn\\
     & = & m^2(\Delta - \ln { m^2\over 4\pi^2 f_{\pi}^2}).
\ee
The divergent part is given by
\be
\Delta & = & {2\over \epsilon}-\gamma -\ln\pi + 1,
\ee
where $\gamma$ is the Euler number and $\epsilon=4-n$.
The decay rate can be evaluated analytically\cite{bijn92} and it reads as
\begin{equation}
\Gamma(\kp) = \nn\\
{\alpha^2 G_8^2 m_K^5 (1-2\sin^2 \theta_W)^2
\over 2^{19}\pi^5 M_Z^4 \sin^4\theta_W \cos^4 \theta_W}
(1-8r_\pi+8r_\pi^3-r_\pi^4 -12r^2_\pi \ln r_\pi)|A(m_K^2)|^2,
\end{equation}
where
\be
r_\pi & = & {m_\pi^2 \over m_K^2}.
\ee
The long distance contribution gives arise to the branching ratio
\be
Br(\kp)_{L.D.}
 & = & 7.71 \times 10^{-18}
\ee
which is roughly of order $10^{-7}$ smaller than that of the short distance
contribution \cite{sd}.

In summary, the $\kp$ amplitude is shown to be much smaller than what has been
calculated before, the effect starts to appears only at $O(P^4)$. This is a
general feature shared by many other GIM
suppressed processes, which will be elaborated fully elsewhere.
With the result of the present work, the uncertainty of the determination of
the standard model parameters from experiments is further restricted and it
makes the proposed experiments more interesting.

\acknowledgments
This work is supported by the National Science Council of the
ROC under contract number NSC84-2122-M-007-013 and NSC84-2112-M-007-041.
One of us (YCL) likes to thank A. Pich for useful conversations.

\newpage

\figure{{\bf Fig 1:} Feynman diagrams which contribute to the tree level
$\kp$ amplitude. The cross stands for the weak interaction. Only diagram
(a) could in principle contribute in the diagonalized basis.
\label{fig1}}


\begin{references}
\bibitem[1]{inam81} T. Inami and C. S. Lim, Prog. of Theor. Phys. {\bf 65}, 297
(1981).

\bibitem[2]{rein89} D. Rein and L. M. Sehgal, Phys. Rev.,
{\bf D39}, 3325 (1989).

\bibitem[3]{hage89} J. S. Hagelin and L. S. Littenberg, Prog. Part. Nucl. Phys.
{\bf 23}, 1 (1989).

\bibitem[4]{lu94} M. Lu and M. Wise, Phys. Lett.
{\bf B324}, 461 (1994).

\bibitem[5]{litt89} L. S. Littenberg, Phys. Rev. {\bf D39} 3322 (1989).

\bibitem[6]{bela91} G. B\'elanger and C. Q. Geng, Phys. Rev. {\bf D43}
140 (1991).

\bibitem[7]{pich95} A. Pich, hep-ph/9502366.

\bibitem[8]{ecke88} G. Ecker, A. Pich and E. de Rafel, Nucl. Phys. {\bf B303},
665 (1988).

\bibitem[9]{kamb90} J. Kambor, J. Missimer and D. Wyler, Nucl. Phys. {\bf
B346}, 17 (1990).

\bibitem[10]{bijn92}
J. Bijnens, G. Ecker and J. Gasser, {\em CERN Report No.} CERN-TH-6504/92.

\bibitem[11]{sd}
Cf. C. Q. Geng, I. J. Hsu, and Y. C. Lin,
Phys. Rev. {\bf D50}, 5744 (1994)

\end{references}
\end{document}